\documentclass[reprint,superscriptaddress,showpacs,amsmath,amssymb,aps,prb]{revtex4-2}

\usepackage{graphicx,subfigure,bbm}% Include figure files
\usepackage{dcolumn}% Align table columns on decimal point
\usepackage{bm}% bold math
\usepackage{amsmath}
\usepackage{amsthm}

\usepackage[english]{babel}
\usepackage[utf8]{inputenc}
\usepackage{graphicx}
\usepackage[colorinlistoftodos]{todonotes}
\usepackage{bbold }
\usepackage{physics}
\usepackage[utf8]{inputenc}
\usepackage{amssymb }
\usepackage{graphicx}
\DeclareGraphicsExtensions{.eps}
\usepackage{epstopdf}
\usepackage{latexsym}
\usepackage{verbatim}

\usepackage{units}
\usepackage{bm}
\usepackage{verbatim}
\usepackage[colorlinks = true,
            linkcolor = red,
            urlcolor  = blue,
            citecolor = magenta,
            anchorcolor = red]{hyperref}
\usepackage{esint}
\usepackage{soul}
\usepackage{cancel}
\usepackage[normalem]{ulem}
\usepackage{empheq}
\usepackage{dsfont}
\DeclareSymbolFont{matha}{OML}{txmi}{m}{it}% txfonts
\DeclareMathSymbol{\varv}{\mathord}{matha}{118}

\newcommand{\sign}{\text{sign}}

\definecolor{GB-color}{RGB}{255,0,255}
\definecolor{AB-color}{RGB}{0,0,255}

\begin{document}
\title{Topological Josephson Junctions in the Integer Quantum Hall Regime}

\author{Gianmichele Blasi}
%\email{gianmichele.blasi@unige.ch} 
\affiliation{Department of Applied Physics, University of Geneva, 1211 Genève, Switzerland}
\author{G\'eraldine  Haack}
\affiliation{Department of Applied Physics, University of Geneva, 1211 Genève, Switzerland}
\author{Vittorio Giovannetti}
\affiliation{NEST, Scuola Normale Superiore and Instituto Nanoscienze-CNR, I-56126, Pisa, Italy}
\author{Fabio Taddei}
\affiliation{NEST, Scuola Normale Superiore and Instituto Nanoscienze-CNR, I-56126, Pisa, Italy}
\author{Alessandro Braggio}
\affiliation{NEST, Scuola Normale Superiore and Instituto Nanoscienze-CNR, I-56126, Pisa, Italy}

\begin{abstract}
Robust and tunable topological Josephson junctions (TJJs) are highly desirable platforms for investigating the anomalous Josephson effect and topological quantum computation applications. Experimental demonstrations have been done in hybrid superconducting-two dimensional topological insulator (2DTI) platforms, sensitive to magnetic disorder and interactions with phonons and other electrons. In this work, we propose a robust and electrostatically tunable TJJ by combining the physics of the integer quantum Hall (IQH) regime and of superconductors. We provide analytical insights about the corresponding Andreev bound state spectrum, the Josephson current and the anomalous current. We demonstrate the existence of protected zero-energy crossings, that can be controlled through electrostatic external gates. This electrostatic tunability has a direct advantage to compensate for non-ideal interfaces and undesirable reflections that may occur in any realistic samples. TJJs in the IQH regime could be realized in graphene and other 2D materials. They are of particular relevance towards scalable and robust Andreev-qubit platforms, and also for efficient phase batteries.
\end{abstract}

\maketitle

\section{Introduction}
Topological Josephson Junctions (TJJ) were proposed in the seminal work of Fu and Kane~\cite{Fu2009a}, merging Josephson physics with topologically-protected helical edge states. These
TJJ have been suggested to host Majorana bound states~\cite{Hegde2020b}, envisioning applications in fault-tolerant quantum computers~\cite{Nayak2008a,Alicea2011,Lahtinen2017,Field2018}, but also intriguing quantum devices for phase-coherent caloritronics~\cite{Bours2019,Scharf2020,Hwang2020,Blasi2020b,Gresta2021,Blasi2021}.
Experimental realizations of TJJ have been proposed by placing two superconducting electrodes on top of topological insulators~\cite{Hart2014,Pribiag2015,Wiedenmann2016,Bocquillon2017,Mandal2022}. Two dimensional topological insulators (2DTIs) are characterized by counter-propagating spin polarized (helical) edge states protected under time-reversal symmetry (TRS)~\cite{Konig2007a,Sinova2015}. 
 However, these topological states are known to be fragile quantum states of matter, suffering from magnetic disorder~\cite{Konig2007a}, Coulomb impurities~\cite{Huang2021}, electron-electron interaction~\cite{Novelli2019}, or electron-phonon interactions~\cite{Budich2012}.
In contrast, integer quantum Hall (IQH) systems with chiral edge states, are by their very fundamental nature very resilient to the  presence of impurities, random electrostatic potentials or interactions~\cite{Halperin1982,Buttiker1988,Mani2016}. Of particular interest for this work, recent experimental achievements have demonstrated the capability of combining QH systems with superconducting contacts~\cite{Takayanagi1998,Moore1999,Hoppe2000,Giazotto2005,Eroms2005,Rickhaus2012,Popinciuc2012,Lee2017,Park2017,Sahu2018,Zhi2019,Zhao2020,Sahu2021,Hatefipour2022,Vignaud2023}.
Experimental realizations of quantum Hall-based Josephson junctions have been reported mainly using graphene \cite{Komatsu2012,Calado2015,Amet2016,Draelos2018,Seredinski2019,Indolese2020} and semiconductors \cite{Guiducci2019,Zhi2019}.

\begin{figure}[!htb]
	\centering
	\includegraphics[width=1\columnwidth]{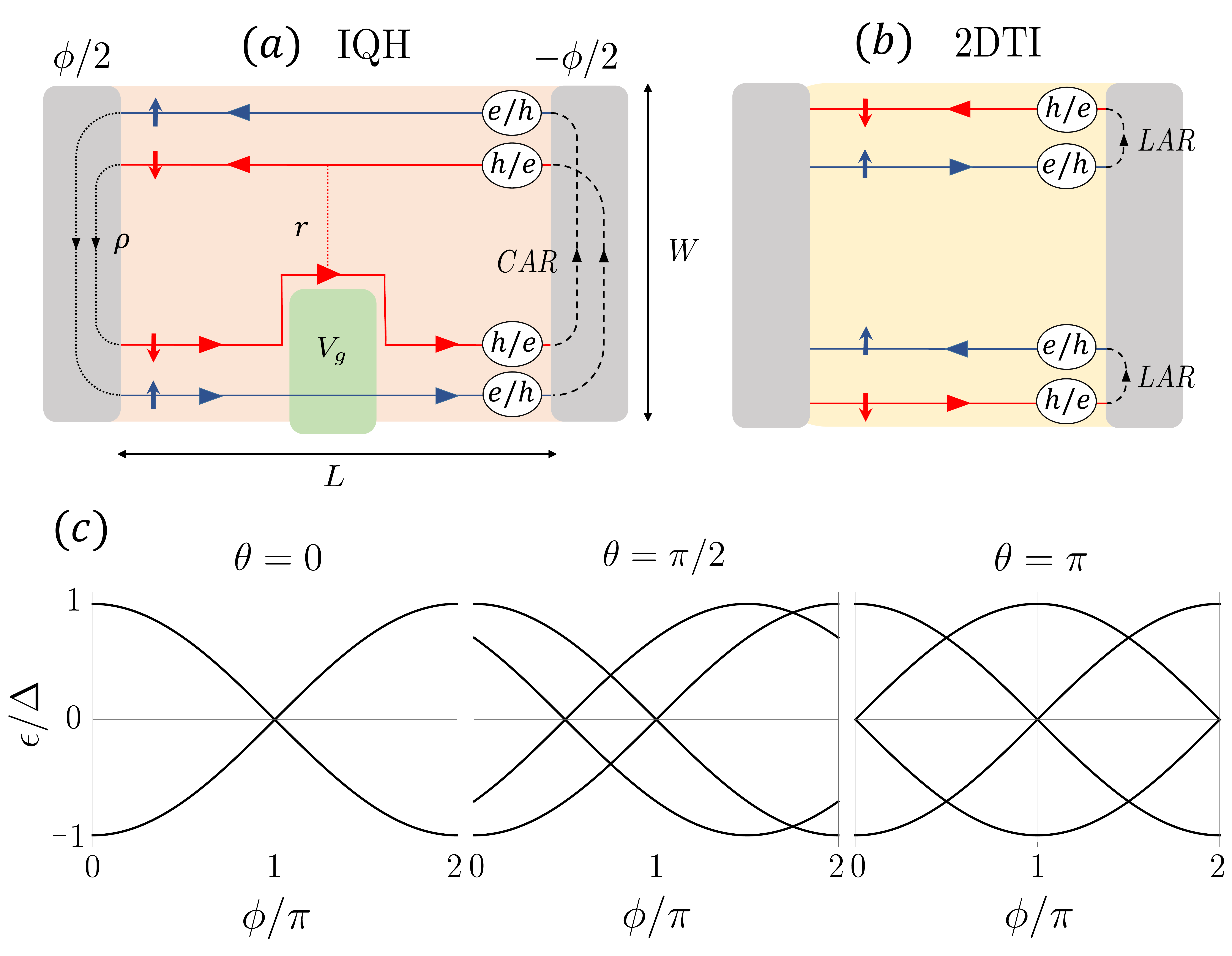}
 	\caption{Sketch of IQH TJJ proposal and associated ABS spectrum. $(a)$ IQH bar of dimensions $W \times L$ at filling factor $\nu=2$, coupled to two superconducting electrodes with phase difference $\phi$. Opposite spin-polarized chiral upper and lower edge states are coupled through CAR processes. Non ideal interfaces allowing for normal reflections are characterized by an amplitude $\rho$. The gate voltage $V_g$ electrostatically controls the phase difference $\theta$ between the spin-polarized right-propagating edge states. A tunnelling amplitude $r$ allows a coupling between right and left propagating edge states. $(b)$ Model for 2DTI TJJ, based on LAR processes between opposite counter-propagating spin-polarized edge states. $(c)$ ABS spectrum as a function of $\phi$ at different $\theta = \{0, \pi/2, \pi\}$ in the IQH TJJ for $r=0$ and $\rho=0$. For $\theta =0$, both 2DTI and IQH TJJ feature exactly the same doubly-degenerate ABS spectrum. $\theta \neq 0$ lifts the degeneracy in the IQH TJJ spectrum, demonstrating full electrostatic control of the topological properties of the IQH TJJ.}
	\label{setup}
\end{figure}

In this work, we propose a TJJ in the IQH regime, with a Hall bar at filling factor $\nu=2$, coupled to two superconductors with gap order parameter $\Delta$. 
For perfect Andreev reflections at the boundaries, we demonstrate that this TJJ in the IQH regime features exactly the same doubly-degenerate Andreev bound state (ABS) spectrum as a 2DTI TJJ with protected zero-energy crossings.
We discuss the mechanisms giving rise to this specific ABS spectrum and show that they are fundamentally different for IQH and 2DTI TJJs. Whereas 2DTI TJJ ABS spectrum results from local Andreev reflection processes (LAR) at the boundaries with a single edge, IQH TJJ ABS spectrum originates from crossed Andreev reflections (CAR) at the boundaries between the two edges.     

In contrast to 2DTI TJJ, our proposal in the IQH is expected to be more robust to magnetic disorder, electron-electron and electron-phonon interactions, and allows for electrostatic control of the ABS spectrum and its zero-energy crossings. 
%In contrast to 2DTI TJJ, our proposal in the IQH is expected to be more robust, and allows for electrostatic control of the ABS spectrum and its zero-energy crossings.
This platform can be used to investigate time-reversal protected physics in cleaner conditions determined by 2DEG/graphene-based quantum Hall physics.
Remarkably, we show that the IQH TJJ exhibits a Josephson current tunable through the gate voltage, allowing to control anomalous current at $\phi=0$. 
These features are promising to implement scalable architectures with multiple Andreev qubits~\cite{Zazunov2003,Chtchelkatchev2003}, as compared to break junctions~\cite{Janvier2015}, nanowire-based setups~\cite{Tosi2019, MatuteCanadas2022} or quantum dot systems~\cite{Oriekhov2021, Kurilovich2021, Pavesic2022}.\\

\section{Model for an IQH TJJ}
Our proposal for an IQH TJJ consists of an integer quantum Hall bar with filling factor $\nu = 2$, coupled to two s-wave superconducting electrodes with phase difference $\phi$~\cite{Stone2011}.
We use a free-fermion theory  as it is usually appropriate for the transport properties in quantum Hall bars, see also App.~\ref{appendix_scattering_matrix}.
The IQH bar with $\nu=2$, due to a finite Zeeman term, can be
characterized by two chiral spin-polarized edge states, propagating spin-up (blue) and spin-down (red) on each edge, see Fig.~\ref{setup} $(a)$. 
The only way an incident electron (hole) can be Andreev-reflected as a hole (electron) with opposite spin is by involving both edges of the Hall bar, as depicted at the interface with the right superconductor in Fig.~\ref{setup}~$(a)$ (equivalent processes involve also the left interface). This mechanism is known as \textit{crossed} Andreev reflection (CAR)~\cite{Ronetti2022} and corresponds to a non-local injection/absorption of a Cooper pair from the superconductors to the weak-link. 
In contrast, the helicity of the edge states in 2DTI TJJs imposes excitations with opposite spin to move in opposite directions on each side of the weak-link.
Therefore, at the interfaces with the superconducting regions, each pair of helical states are coupled through \textit{local} Andreev reflection (LAR) processes, involving degrees of freedom of a single edge. 
Hence, upper and lower edges can be considered as two independent sectors, see Fig.~\ref{setup} $(b)$.

One of the key advantages of the IQH TJJ with respect to the 2DTI TJJ is its tunability via a top voltage gate $V_g$ (in green on the sketch). 
In this setup $V_g$ controls the filling factor of the region underneath the gate, modifying the path of the inner edge state. In particular, when the filling factor induced by $V_g$ is equal to 1, the inner edge state has to propagate around the border of the gate. As a consequence, particles propagating along this modified edge state acquire a $V_g$-controlled dynamical phase $\theta=k_F \Delta L$ proportional to the length difference $\Delta L$ between the outer and inner edges.  
Here we assume, for simplicity, that the Fermi velocity is the same for both spin-up and spin-down particles. Any difference in velocity would cause particles to acquire an effective dynamical phase (which can also depend on the magnetic flux), even without the gate. Without loss of generality, this additional dynamical phase could be included in $\theta$.

As an additional feature, it is important to note that the top gate, as it penetrates the sample, also enables the spin-down channel of the lower edge to approach the upper edge. This permits tunneling to occur between the opposite sides of the Hall bar.
In our model this process is described by a tunneling amplitude $r$. If $\abs{r}^2=0$ no tunneling occurs, while if $\abs{r}^2=1$ spin-down excitations are completely back-reflected on the opposite edge thus destroying the Josephson coupling between superconductors~\cite{Vigliotti2022}.

In realistic samples, ordinary reflections occurring at the interfaces with the superconductors between the upper and lower edges may take place. 
For example, an electron (or hole) in the upper edge may reach coherently the opposite edge as indicated in Fig.~\ref{setup}~$(a)$ by black dotted lines at the interface with the left superconductor. 
These processes are in direct competition with the CAR processes operating at the same interface.
To investigate this aspect, we account for normal reflections at the left and right interfaces with reflection amplitudes $\rho_L, \rho_R$.
When $\abs{\rho_{L,R}}^2 = 0$, no ordinary reflections can occur and particles can only undergo perfect CAR, whereas $\abs{\rho_{L,R}}^2=1$ corresponds to only ordinary reflections taking place, completely spoiling the Josephson coupling.
Interestingly, we show below that left-right symmetry in the device, corresponding to $\rho\equiv \rho_L = \rho_R$, is sufficient to protect zero-energy crossings in the ABS spectrum even in the presence of normal reflections.\\

\begin{figure*}
  \includegraphics[width=0.8\textwidth]{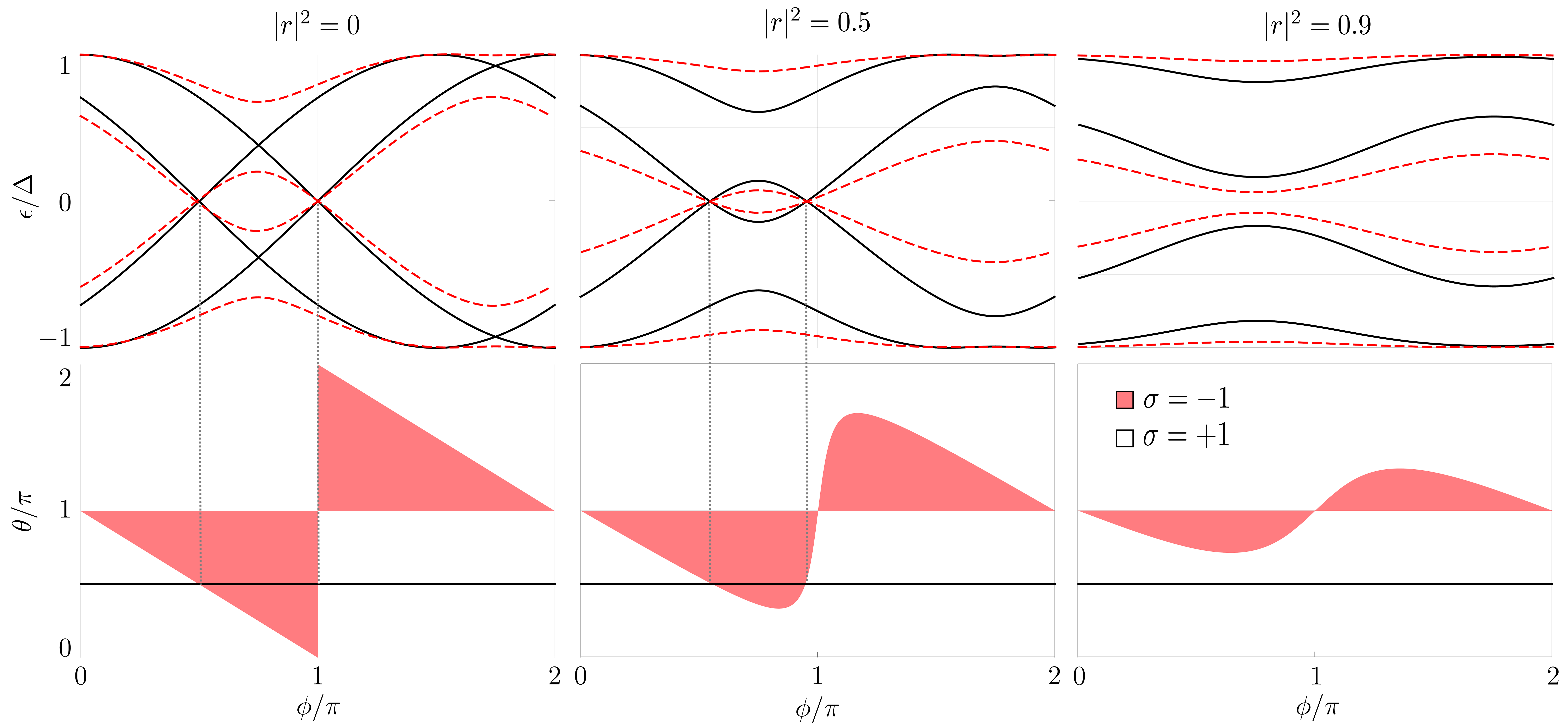}
  \caption{ABS spectrum (top panels) as a function of the superconducting phase difference $\phi$ and color plot of the figure of merit $\sigma$ (bottom panels) in the $\phi-\theta$ plane. From left to right panels, tunneling probabilities from the lower edge to the upper edge is varied, $\abs{r}^2 = \{0,0.5, 0.9\}$.
Top panels: ABS spectra as a function of $\phi$ obtained for $\abs{\rho}^2=0$ (black solid curves), and $\abs{\rho}^2=0.5$ (dashed red curves). Notice that the positions of zero-energy crossings are the same for dashed and solid curves (they do not change with $\rho$).
As tunneling amplitude $r$ is increased, Josephson coupling decreases, leading to a gapped spectrum for $\abs{r}^2=0.9$. 
Bottom panels: red regions represent $\sigma=-1$, white regions indicate $\sigma=+1$. Black horizontal thick lines correspond to $\theta=\pi/2$, used to plot the ABS spectra. }
	\label{ABS2}
\end{figure*}

\section{ABS spectrum} 
\label{sec_ABS_spectrum}
We derive and analyze the properties of the ABS spectrum of the IQH TJJ within a scattering matrix (S-matrix) approach.
The IQH bar is described as a normal metal weak link by a block-diagonal S-matrix $S_{N}$ in the electron-hole subspace, whereas CAR processes at the interfaces with the superconducting electrodes are captured by an off-diagonal S-matrix $S_A$. Exact forms of these S-matrices are provided in the Appendix \ref{appendix_scattering_matrix}. 
% The ABS spectrum corresponds to the energies $\epsilon_p$, solutions of the self-consistent secular problem $\text{Det}\left[1-\alpha(\epsilon_p)S_AS_N\right]=0$, with $\alpha(\epsilon_p)=\exp{i\left[\frac{\epsilon_p}{\Delta}\frac{L}{\xi}-\arccos{(\frac{\epsilon_p}{\Delta})}\right]}$ being the energy-dependent Andreev reflection factor~\cite{Beenakker1992a}. 
The ABS spectrum corresponds to the energies $\epsilon_p$, which are solutions of the self-consistent secular problem given by:
\begin{equation}
    \text{Det}\left[1-\alpha(\epsilon_p)S_A\exp\left(i\left[\frac{\epsilon_p}{\Delta}\frac{L}{\xi}\mathds{1}+\frac{\pi\Phi}{\Phi_0}\tau_z\right]\right)S_N\right]=0 .
\end{equation}
Here, $\alpha(\epsilon_p)=\exp\left(-i\arccos\left(\frac{\epsilon_p}{\Delta}\right)\right)$ represents the energy-dependent Andreev reflection factor~\cite{Beenakker1992a}. The term $\exp\left(i\left[\frac{\epsilon_p}{\Delta}\frac{L}{\xi}\mathds{1}+\frac{\pi\Phi}{\Phi_0}\tau_z\right]\right)$ corresponds to the phase acquired by particles due to the finite length of the junction, which also takes into account the presence of the magnetic flux $\Phi=LWB$ ($\Phi_0=h/2e$ represents the flux quantum)~\cite{Tkachov2015}. Here $\mathds{1}$ and $\tau_z$ represent the identity and $z$-Pauli matrices in the particle-hole space.
However, it is  important to note that this term equals 1 in the limit of a short junction, i.e., when $L\ll \xi$.
% We emphasize that in the analysis that follows  we will assume a constant magnetic field, but not investigate the effect of a modulation of the magnetic-flux through the junction. In the latter case one would need to include other effects such as the change of the filling factor or  not-universal effects occurring at the interfaces with the superconductors.
We would like to emphasize that in the subsequent analysis, we will assume a fixed magnetic field, without investigating the effect of a modulation of the magnetic flux through the junction. In the latter case, one would need to consider other effects, such as the change of the filling factor or non-universal effects occurring at the interfaces with the superconductors.
% Here t
The index $p=\{1,2,\cdots\}$ labels the different ABS discrete solutions. $L$ is the length of the IQH weak-link and $\xi=\hbar v_F/\Delta$ is the superconducting coherence length with $\Delta$ being the superconducting gap. In the limit of a short junction ($L\ll\xi$), with no ordinary reflections at the interfaces ($\rho=0$) and no tunneling ($r=0$), the energies $\epsilon_p$ are found analytically (see Appendix \ref{app_ABS_analitical} for more details) and are shown in Fig.~\ref{setup} $(c)$.  
In case $V_g = 0$, no phase difference $\theta$ is present, and one recovers the doubly degenerate ABS spectrum of a 2DTI TJJ with $\epsilon_p = \pm\Delta \abs{\cos(\phi/2)}$ \cite{Fu2009a,Beenakker2013}. 
Figure~\ref{setup} $(c)$ clearly shows the zero-energy crossing at $\phi = \pi$ for $\theta =0$. 
Notice that the ABS spectrum of the IQH TJJ is doubly degenerate. It corresponds to the ABS spectrum of the two independent copies of the ABS spectra of the two edges of the 2DTI TJJ sketched in Fig.~\ref{setup} $(b)$. 
This degeneracy can be lifted by tuning the dynamical phase $\theta$ controlled by $V_g$.
As shown in the central panel of Fig.~\ref{setup}~$(c)$, for $\theta = \pi/2$,  one of the two copies of the spectrum gets rigidly shifted to the left, and the ABS spectrum presents two zero-energy crossing, at $\phi=\pi$ and at $\phi=\pi/2$.
Importantly, such an electrostatic control can not be easily implemented in a 2DTI TJJ. 
%The rightmost panel in Fig.~\ref{setup}~$(c)$ refers to the case $\theta=\pi$ in which one of the branches is shifted to the left by a phase $\phi=\pi$.

At this stage, we have shown that fundamentally different processes underlie the physics of the 2DTI TJJ and of the IQH TJJ: LAR versus CAR processes, presence or not of TRS, and helical versus chiral nature of the edge states, respectively. Hence, it is truly remarkable that our proposal of an IQH TJJ exhibits the same ABS spectrum as the 2DTI TJJ, with an additional tuning parameter $\theta$.

\section{Protected zero-energy crossings} 
\label{sec_protection}
We now investigate the protection of the zero-energy crossings of the IQH TJJ, in presence of ordinary reflections ($\rho_L, \rho_R \neq 0$) and in presence of finite tunneling $r$. 
We first consider the case $\rho_L = \rho_R \equiv \rho$. For $\rho \neq 0$, our numerical calculations show that energy gaps open at finite energy, while ABS crossings at zero energy remain unchanged. 
This is illustrated in top panels of Fig.~\ref{ABS2} where we represent the ABS spectrum as a function of $\phi$ for $\abs{\rho}^2=0$ (black solid curves) and $\abs{\rho}^2=0.5$ (red dashed curves). 
These results indicate that zero-energy crossings in the ABS spectrum are protected against perturbations which do not break left/right symmetry in the system, i.e.~when $\rho_L=\rho_R$. We further verified that zero-energy crossings at a fixed $\theta$ disappear when $\rho_L\neq \rho_R$, even if they only differ by a phase. Interestingly, the electrostatic parameter $\theta$ specific to the IQH TJJ can be tuned to ensure the presence of zero-energy crossings for any complex value of $\rho_L$ and $\rho_R$ (see Appendix \ref{app_Asymmetric_junction} for more details).

We then observe that the figure of merit $\sigma$ - introduced in \cite{Beenakker2013} for a 2DTI TJJ to gauge the parity of the ABS spectrum - also allows us to single out remarkable protection features of the zero-energy crossings of the IQH TJJ ABS spectrum in presence of finite tunneling $r$. 
This figure of merit $\sigma=\pm$ can be computed analytically for our model and is shown to not depend on $\rho$ (see Appendix \ref{app_ABS_analitical} for the definition and details):
\begin{eqnarray}
\label{sign_Pf}
\sigma&=&\sign{\left\{ \cos ^{-1}\left[-\frac{\sqrt{1-\abs{r}^2} \sin (\phi )}{\sqrt{-2 \sqrt{1-\abs{r}^2} \cos (\phi )-\abs{r}^2+2}}\right]\right\}} \,. \nonumber \\
&&
\end{eqnarray}

Bottom panels in Fig.~\ref{ABS2} show $\sigma$ as a function of the superconducting phase difference $\phi$ and of the dynamical phase $\theta$, for different values of the tunneling probability ($\abs{r}^2= \{0, 0.5, 0.9\}$ from left to right). 
The black horizontal thick lines correspond to $\theta=\pi/2$, used to plot the corresponding ABS spectra (top panels).
Zero-energy crossings correspond to the values of $\phi$ in the region plot for which the horizontal black line intercepts the boundaries at which $\sigma$ changes its sign (see vertical dotted lines). Again, these zero-energy crossings do not change with $\rho$ for fixed $r$. The figure also indicates that the distance between the zero-energy crossings reduces when increasing $\abs{r}^2$, until the zero-energy crossings disappear and are replaced by a gap at zero energy, see for instance the last panel for $\vert r \vert^2 = 0.9$. No crossing occurs at $\theta = \pi/2$, but changing the electrostatic parameter $\theta$ would allow one to recover zero-energy crossings.
Hence, it is possible to tune $\theta$ to compensate for higher $r$, and therefore to protect the zero-energy crossings over a wider range of tunneling amplitudes. In particular, for $\theta = \pi$, the two zero-energy crossings always occur at $\phi=0$ and $\phi = \pi$, independently of $\abs{r}^2$ if $\abs{r}^2\neq 1$. The electrostatic control of the IQH TJJ make the zero-energy crossings robust against the presence of back-scattering within the IQH bar, as long as the Josephson coupling is not fully suppressed, i.e.~as long as $\abs{r}^2< 1$. 

It is important to emphasize that we have assumed the short junction limit, where $L\ll\xi$. This allowed us to derive analytical results for the ABS spectrum (Sec. \ref{sec_ABS_spectrum}), compare the behavior of the zero-energy crossing of the ABS spectrum with the parity figure of merit (Sec. \ref{sec_protection}), and obtain analytical expressions for the Current Phase Relationship (CPR) and critical current (as demonstrated in the next Sec. \ref{sec_CPR_critical}).
However, while the assumption of a short junction is a strong one for the experimental implementation of the system, we have numerically confirmed that the zero energy crossing of the lower energy ABS does not change with respect to the short-junction case, regardless of the value of $L$.
% \blue{the presence of the zero energy crossings of the lower energy ABS persist (i.e. they cannot be removed) regardless of the value of $L$ and of the magnetic flux $\Phi$.}
In this sense, we assert that the topological protection of such crossings is not specifically tied to the short-junction limit.

\section{Josephson current}
\label{sec_CPR_critical}
Characterizing the Josephson current as a function of the parameters of the setup is key to access the property of the IQH TJJ.
In IQH systems it has been measured in Refs. \cite{Komatsu2012,Popinciuc2012,Calado2015,Amet2016,Draelos2018,Zhi2019,Zhi2019,Seredinski2019,Guiducci2019,Indolese2020}.
\begin{figure}[htb]
	\centering
	\includegraphics[width=1.0\columnwidth]{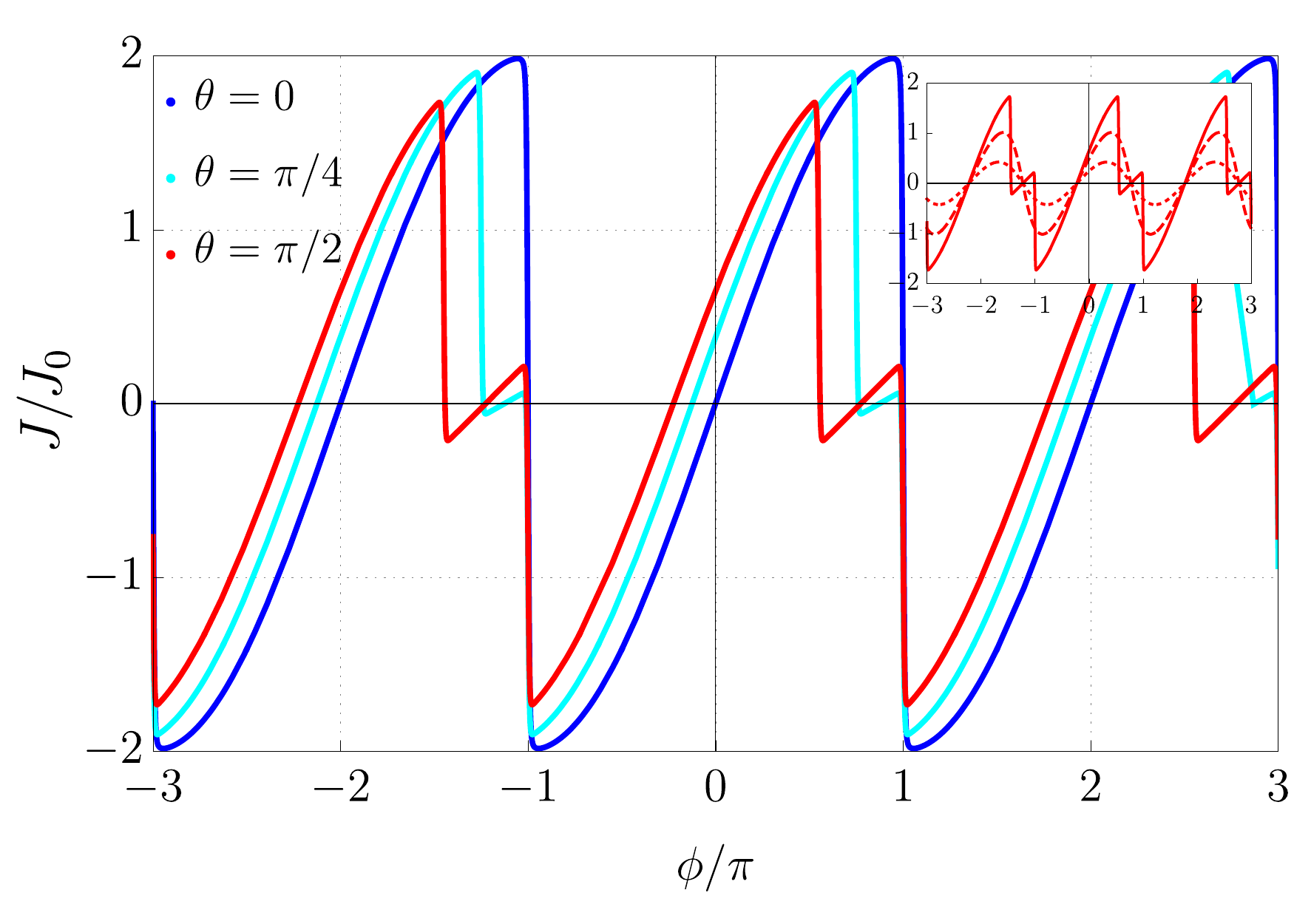}
 	\caption{Current-phase relationship of the IQH TJJ for $r=0$ and $\rho=0$ (main panel), in units of $J_0 = 4e\Delta/\hbar$. The different colored lines corresponds to different gate voltages which change the $\theta$ values as reported in the label. In the inset we show the effect of different values of tunnelling $\abs{r}^2=0.1,0.5,0.9$ (solid, dashed and dotted line, respectively)  for a given value of $\theta=\pi/2$. The temperature is chosen to be very low, namely $k_BT\ll \Delta$.}
	\label{CPRPlot}
\end{figure}

\begin{figure}[htb]
	\centering
	\includegraphics[width=0.9\columnwidth]{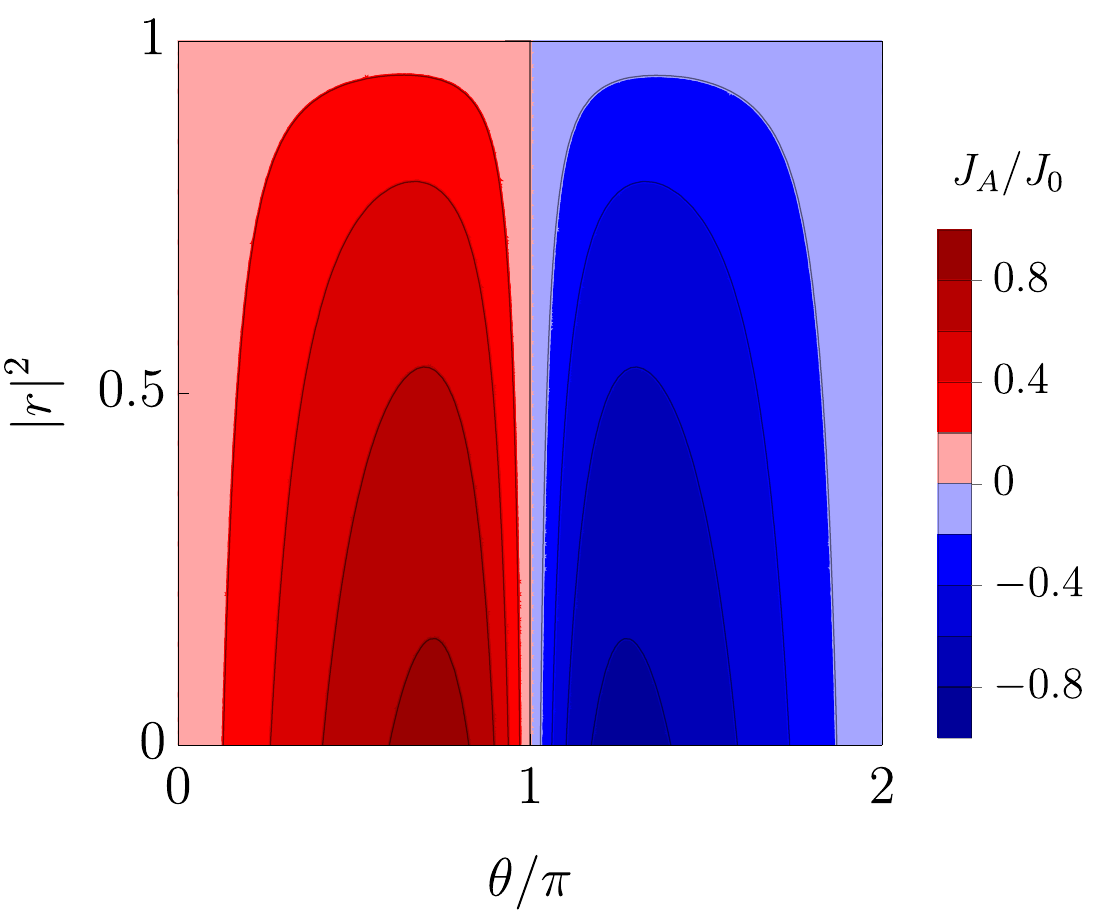}
 	\caption{Anomalous Current $J_A$ in units of $J_0=\frac{4e\Delta}{\hbar}$ as function of $\theta$ and the tunneling probability $\abs{r}^2$ for $\rho=0$.}
	\label{JA}
\end{figure}

Assuming short junctions ($L\ll \xi$), we recall that the Josephson current $J(\phi)$, or current phase relation (CPR), is given by \cite{Beenakker1992a,Ma1993}: $J (\phi)=-\frac{2e}{\hbar}\sum_{p}\tanh\qty(\epsilon_p/2k_BT)\frac{d\epsilon_p}{d\phi},$ where $T$ is the temperature and $\epsilon_p$ are the ABS energies. 
For the IQH TJJ, we obtained an exact expression of $J(\phi)$ in the limit of  negligible normal reflections (i.e.~$\abs{\rho}^2=0$). This provides analytical insights on the role of the different parameters of the system, in particular $\theta$ and $r$:
\begin{align}
\label{CPR}
J(\phi)=&-\frac{J_0\sqrt{1-\abs{r}^2}\sin{(\frac{\theta}{2}+\phi)}}{\sqrt{\abs{r}^2\cos{(\frac{\theta}{2}+\phi)}^2+\sin{(\frac{\theta}{2}+\phi)}^2}}\times\nonumber\\
&\sum_{\gamma=\pm}\gamma\cos{\left[\Gamma_{\gamma}\left(\phi,\theta\right)\right]}\tanh{\left\{\frac{\Delta}{2k_BT}\sin{\left[\Gamma_{\gamma}\left(\phi,\theta\right)\right]}\right\}}
\end{align}
with $\Gamma_{\gamma}=\frac{\theta}{4}+\gamma\frac{\pi}{4}+\gamma\tan^{-1}{\left[\frac{\sqrt{\abs{r}^2}\sin{(\frac{\theta}{2}+\phi)}}{\sqrt{\abs{r}^2\cos{(\frac{\theta}{2}+\phi)}^2+\sin{(\frac{\theta}{2}+\phi)}^2}}\right]}$,
and $J_0=\frac{4e\Delta}{\hbar}$. Zero-energy crossings in the ABS spectrum correspond to abrupt jumps in the Josephson current, clearly shown in Fig.~\ref{CPRPlot} (see Refs.~\cite{Dolcini2015, Marra2016, Pientka2017} for the equivalence in 2DTI TJJ). In analogy to the ABS spectrum, the Josephson current can be controlled by the voltage gate $V_g$, which acts similarly to the manipulation of spin-orbit coupling in Refs.~\cite{Marra2016,Ehren2016,Pientka2017,Arrachea2019,Chaitanya2020}. Let us note that the presence of back-scattering within the IQH bar ($\abs{r}^2 \neq 0$) smooths the CPR behavior in a comparable way to temperature or other mechanisms of losses~\cite{Blasi2019a}, as shown in the inset of Fig.~\ref{CPRPlot}.

Remarkably, the CPR also presents a finite anomalous component at $\phi=0$, $J_A\equiv J\left(\phi = 0\right)$. We show its behaviour as a function of the phase difference $\theta$ controlled by $V_g$ and as a function of back-scattering (parameter $\vert r \vert^2$) in Fig.~\ref{JA}. It is maximal for $\abs{r}^2=0$ and vanishes when the Josephson coupling disappears ($\abs{r}^2=1$). We note that $J_A$ is a $2\pi$-periodic odd function of $\theta$, i.e.~ $J_A(\theta)=-J_A(-\theta)$, implying that $J_A = 0$ for $\theta = \pi$ mod[$\pi$].
This anomalous component is a common feature of topological Josephson junctions~\cite{Tanaka2009,Dolcini2015,Marra2016}, also discussed in presence of magnetic fields or ferromagnetic exchange \cite{Ostaay2011,Szombati2016,Alidoust2017,Assouline2019,Mayer2020}. This property opens the way to exploit the IQH TJJ as a gate controlled phase battery without engineering complex interplay between spin-orbit and magnetic field. Phase batteries were envisioned and recently demonstrated experimentally to efficiently provide a constant phase difference between two superconductors in a quantum circuit~\cite{Linder2015, Pal2019, Strambini2020}. This should highly increase the scalability and tunability of the device.

Another important quantity which characterizes the IQH TJJ system is represented by the critical current defined as $J_C=\text{max}_{\phi}\left\{\abs{J(\phi)}\right\}$. 
Starting from Eq.~\eqref{CPR}, and considering the limit of high temperature (i.e.~ $k_BT\gg\Delta$)~\footnote{Note that the gap $\Delta$ induced on the edge states is usually smaller than the intrinsic gap of the superconducting leads. Therefore superconductivity in the leads is not suppressed when $k_BT\gg\Delta$.}, the critical current takes the form $J_C\simeq J_0 \frac{\Delta}{2 k_BT}\abs{\cos{(\theta/2)}}$.
This coincides with the result obtained in Ref. \cite{Blasi2019a} where the critical current has been used as a signature of  entanglement symmetry manipulation of Cooper pairs (CPs) in the condensate exchanged between the superconducting leads.
The critical current $J_C$ is of particular relevance as it not only reflects the strength of the Josephson coupling, but it also has a fundamental role in fixing the minimal kinetic Josephson junction inductance. It is therefore relevant in the design of dispersive and resonant measurement setups, for instance in the context of Andreev qubits. 
Andreev qubits were first proposed in Refs.~\cite{Zazunov2003,Chtchelkatchev2003}, and are of particular interest for applications in quantum information~\cite{Bretheau2013,Hays2018,Hays2021}.
Recently, in the context of Andreev qubits, nanowires-based Josephson junctions with strong spin-orbit coupling have been proposed and investigated~\cite{Tosi2019,MatuteCanadas2022}, as well as quantum dot systems~\cite{Oriekhov2021,Kurilovich2021,Pavesic2022}. In Appendix \ref{app_Andreev_qbit}, we discuss the microwave excitation spectrum of our IQH TJJ in the realistic case of finite-length weak link ($L=\xi$), which could be used to store and elaborate quantum information. On top of exhibiting key properties for implementing an Andreev qubit similarly to the above mentioned platforms, our setup allows for a much refined electrostatic control of the ABS spectrum. This would be beneficial for the scaling to a multiple qubits platform with respect the atomic break junctions~\cite{Janvier2015} or nanowire-based setups~\cite{Janvier2015,MatuteCanadas2022}. 

\section{Experimetal Aspects}
In this section, we briefly discuss some key aspects of our experimental setup. Previous studies, such as Ref.~\cite{Karmakar2011} and Ref.~\cite{Karmakar2013}, have shown the ability to spatially separate the $\nu=2$ edge modes for spin-up and spin-down at a magnetic field of about $4.5~$T and of about $2.5~$T, respectively. More recently, the authors of Ref.~\cite{Guiducci2019} have examined an InAs quantum well in contact with $\text{Nb}$ superconducting leads and found that the system demonstrates well-developed quantized Hall plateaus at a temperature of $300$ mK and a magnetic field of $3~$T, resulting in a filling factor of $\nu=2$. The electronic density was also controlled using a gate voltage, which allowed a filling factor of $\nu=1$ at a gate voltage of $-1~$V and $\nu=2$ at a gate voltage of $3~$V. These experimental results demonstrate that the proposed setup is realistic with current technology.
\\
Regarding the size of our system, we can estimate the coherence length of the Cooper pairs using the value of the Nb gap ($\Delta_{\text{Nb}}=1.23~$meV) and the Fermi velocity in the edge state ($v_F\approx 10^{6}$m/s) to be $\xi=(\hbar v_F)/(\pi \Delta_{\text{Nb}})\approx200$ nm. It is important to note, however, that the strong magnetic field and imperfect coupling with the quantum Hall edge states may further reduce the effective gap, thus increasing the coherence length. Therefore, assuming a length $L$ of our IQH TJJ of a few hundred nanometers, which is achievable using current lithography techniques, would be sufficient to realize the proposed setup.
\\
Furthermore, regarding the width $W$ of the bar, it is worth noting that it can exceed $\xi$ as the CAR is mediated by the Andreev reflections taking place on the edge states traveling across the entire interface with the superconductors.

\section{Conclusion} 

In this paper we have investigated how to realize an artificial time-reversal-symmetric topological Josephson junction with a paradigmatic time-reversal-broken phase, i.e. the IQH state at $\nu=2$, coupled to superconductors via CAR processes. First, we have demonstrated how the two models map one into the other, showing the emergence of topological protected zero-energy crossings. 
Our IQH platform has the advantage of being much more robust against impurities or other perturbations that destroy the perfect conductance quantization, and provides better prospects in terms of scalability.
Specific to this proposal, the IQH TJJ can be manipulated by electric means and does not require spin-orbit effects or careful manipulation of the magnetic field orientation. We investigate in details how the coupling between upper and lower edge states and the gate voltage control
the ABS spectrum, the protected zero-energy crossings and the Josephson current. 
We also show how an anomalous Josephson current can be generated and controlled by the gate voltage, without requiring complex interplay between spin-orbit coupling and magnetic fields. 
We also discuss the role of other non-idealities such as imperfect interfaces with the superconductors.
Examples with filling factor $\nu=2$ have been demonstrated both in 2DEG samples~\cite{Takagaki1998,Wan2015} and in 2D materials, such as graphene~\cite{Lee2017,Michelsen2022}.   

\section{Acknowledgements} 

GB and GH acknowledge support from the Swiss National Science Foundation through the NCCR QSIT and GH additionnally thanks the SNSF for support through the NCCR SwissMAP and a starting grant PRIMA PR00P2$\_$179748. AB thanks the EU’s Horizon 2020 research and innovation program under Grant Agreement No. 964398 (SUPERGATE) and No. 101057977 (SPECTRUM). AB and FT acknowledge the Royal Society through the International Exchanges between the UK and Italy (Grants No. IEC R2 192166 and IEC R2 212041).

%%%%%%%%%%%%%%%%%%%%%%%%%%%%%%%%%%%%%%%%%%%%%%%%%%%%%%%%%%%%%%%%%%%%%%%%%%%%%%%%%%%%%%%%%%
%%%%%%%%%%%%%%%%%%%%%%%%%%%%%%%%%%%%%%%%%%%%%%%%%%%%%%%%%%%%%%%%%%%%%%%%%%%%%%%%%%%%%%%%%%
%%%%%%%%%%%%%%%%%%%%%%%%%%%%%%%%%%%%%%%%%%%%%%%%%%%%%%%%%%%%%%%%%%%%%%%%%%%%%%%%%%%%%%%%%%

\appendix

\section{Scattering matrices}
\label{appendix_scattering_matrix}
We use a free-fermion theory (the scattering theory) since it is usually appropriate for the transport properties in quantum Hall bars. Indeed,
experiments, see for example Ref.~\cite{Karmakar2011}, show that at equilibrium and in the linear regime, the properties of the system
are well predicted by the non-interacting theory. This is true even in the presence of interchannel tunnelling
or equilibration, where one can expect a possible role of the interactions. In fact, in order to detect an effect of the
interaction between edge channels, one needs typically long edges with lengths of tens micrometres as
discussed in Ref.~\cite{Hashisaka2018}. In other words, our junction is too short to expect to see these effects.

The system is modeled by specifying the scattering matrices (S-matrices) which describe the weak-link $S_N$ and the interfaces with the superconductors $S_A$, that can be written in the electron-hole space as
\begin{align}
	\label{SA_SN}
	S_N=
	\begin{pmatrix}
		s_0&0\\
		0&s_0^*
	\end{pmatrix},
	&&
	S_A=\begin{pmatrix}
		0&r_A\\
		r_A^*&0
	\end{pmatrix}.
\end{align}
Diagonal blocks represent the coupling between electron and electron (or hole and hole), while off-diagonal blocks account for the electron-hole coupling.
In Eq.~\eqref{SA_SN}, the Andreev matrix $r_A$ ($r_A^*$) relates the  holes (electrons) impinging onto the superconductors to the corresponding electrons (holes) going back in the opposite edge of the weak-link and depends on the superconducting phases $\pm\phi/2$ acquired at the left/right superconductors. 
As an example, the S-matrix $r_A^*$ of Eq.~\eqref{SA_SN} which relates impinging electrons with Andreev reflected holes at interfaces with superconductors, takes the following form:
\begin{equation}
	\label{rA*}
	\mqty(b^{\uparrow}_+\\
	b^{\downarrow}_+\\
	b^{\uparrow}_-\\
	b^{\downarrow}_-)=
	\begin{pmatrix}
		\begin{matrix}
			0&e^{i\phi/2}\\
			e^{i\phi/2}&0
		\end{matrix}&\begin{matrix}
			0&0\\
			0&0
		\end{matrix}\\
		\begin{matrix}
			0&0\\
			0&0
		\end{matrix}&\begin{matrix}
			0&e^{-i\phi/2}\\
			e^{-i\phi/2}&0
		\end{matrix}
	\end{pmatrix}_{r_{A}^*}
	\mqty(c^{\uparrow}_-\\
	c^{\downarrow}_-\\
	c^{\uparrow}_+\\
	c^{\downarrow}_+),
\end{equation}
(a similar equation can be written for the S-matrix $r_A$~\cite{Beenakker1992a}).
Here we indicated by $c_\pm^\sigma$ and $b_\pm^\sigma$ the incoming and outgoing electrons and holes respectively, with $\pm$ labeling the direction of propagation of particles ($+$ for right movers, $-$ for left movers) with spin $\sigma=\uparrow,\downarrow$. 

The S-matrix $S_N$, on the other hand, is block-diagonal. Specifically, the block-matrix component $s_0$, which relates incoming and outgoing electrons only, is constructed (see Refs.~\cite{Datta1997,Blasi2022}) from the combination of S-matrices:
\begin{equation}
	\label{s0_comb}
	s_0= s_L \circ s_M \circ s_R.
\end{equation}
Here $s_M$ is the S-matrix which describes the presence of the gate acting only on spin down particles, while $s_{L/R}$ accounts for the possible presence of ordinary reflections at the interfaces with the superconductors.
More specifically, the matrices $s_M$ and $s_{i}$ (with $i=L,R$), introduced in Eq.~\eqref{s0_comb}, can be written in following form\\
\begin{align}
\label{sM_sL_sR}
	\mqty(c^{\uparrow}_-\\
c^{\downarrow}_-\\
c^{\uparrow}_+\\
c^{\downarrow}_+)_{out}&=
\begin{pmatrix}
	\begin{matrix}
		0&0\\
		0&r
	\end{matrix}&\begin{matrix}
		1&0\\
		0&t_{LR}
	\end{matrix}\\
	\begin{matrix}
		1&0\\
		0&t_{RL}
	\end{matrix}&\begin{matrix}
		0&0\\
		0&r
	\end{matrix}
\end{pmatrix}_{s_{M}}
\mqty(c^{\uparrow}_+\\
c^{\downarrow}_+\\
c^{\uparrow}_-\\
c^{\downarrow}_-)_{in}\nonumber\\
	\mqty(c^{\uparrow}_-\\
c^{\downarrow}_-\\
c^{\uparrow}_+\\
c^{\downarrow}_+)_{out}&=
\begin{pmatrix}
	\begin{matrix}
		\rho_i&0\\
		0&\rho_i
	\end{matrix}&\begin{matrix}
		\tau_i&0\\
		0&\tau_i
	\end{matrix}\\
	\begin{matrix}
		\tau_i&0\\
		0&\tau_i
	\end{matrix}&\begin{matrix}
		\rho_i&0\\
		0&\rho_i
	\end{matrix}
\end{pmatrix}_{s_{i}}
\mqty(c^{\uparrow}_+\\
c^{\downarrow}_+\\
c^{\uparrow}_-\\
c^{\downarrow}_-)_{in}
\end{align}
with
\begin{align}
	\label{BC}
	s_M= \begin{cases}
		r &= e^{i\theta/2}\cos{\left(\omega\right)}\\
		t_{RL} &= -e^{i\theta}\sin{\left(\omega\right)}\\
		t_{LR} &= \sin{\left(\omega\right)}
	\end{cases} &&
	s_i= \begin{cases}
		\rho_i &= \cos{\left(\eta_i\right)}\\
		\tau_i &= i \sin{\left(\eta_i\right)}
	\end{cases} 
\end{align}
Here the S-matrix $s_M$ depends on two angle parameters: $\omega\in[0,\pi/2]$ which controls the reflection amplitude $r$ of spin down particles between different edges, and $\theta\in[0,2\pi]$ (controlled by the gate voltage $V_g$) which is the dynamical phase acquired by spin down particles moving around the gated region. 
The S-matrix $s_{i}$ is controlled by the parameter $\eta_{i}\in[0,\pi/2]$ which determines the amplitude $\rho_i$ of ordinary reflections at the interface $i=L,R$. If $\eta_i=\pi/2$, then $\rho_i=0$ so only CAR can occur, while if $\eta_i=0$ no CAR can occur and only ordinary reflections can take place at the interface. 

\section{Analytical expression of ABS and figure of merit $\sigma$}
\label{app_ABS_analitical}

We provide the analytical expressions of the ABS eigenenergies obtained in the limit of a short junction ($L\ll\xi$), with no ordinary reflections at the interfaces ($\rho=0$). The ABS energies have been obtained by solving the self-consistent secular problem~\cite{Beenakker1992a}
\begin{equation}
	\label{secular}
	\text{Det}\left[1-\alpha(\epsilon_p)S_AS_N\right]=0,
\end{equation}
with $S_A$ and $S_N$ defined in Eqs.~\eqref{SA_SN}, and $\alpha(\epsilon_p)=\exp\left(-i\arccos\left(\frac{\epsilon_p}{\Delta}\right)\right)$ being the energy-dependent Andreev reflection factor.
Their explicit expressions take the following form:
\begin{equation}
	\label{eq:energy_sol}
	\epsilon_p=\Delta\abs{\sin{\left(\Xi_{\pm}\right)}}
\end{equation}
with
\begin{equation}
	\label{eq:energy_sol_2}
	\Xi_{\pm}=\frac{\theta}{4}\pm\frac{\pi}{4}\pm\tan^{-1}{\left[\frac{\sqrt{1-\abs{r}^2}\cos{(\frac{\theta}{2}+\phi)}}{\sqrt{\abs{r}^2\sin{(\frac{\theta}{2}+\phi)}^2+\cos{(\frac{\theta}{2}+\phi)}^2}}\right]}.
\end{equation}
Due to particle-hole symmetry, each eigenstate with energy $\epsilon_p$ has a counterpart with energy  -$\epsilon_p$, so for each value of the phase $\phi$, we have four solutions of the ABS energies.
We used Eqs.~\eqref{eq:energy_sol} and \eqref{eq:energy_sol_2} to plot the ABS spectrum in Fig.~1~$(c)$ and Fig.~2 (black solid curves in top panels) in the main text. 
We have also derived an analytical solution of the ABS eigenenergies also for $\rho\neq0$, but the expressions do not bring additional understanding. They are available upon request for interested readers.

The figure of merit $\sigma$, given in Eq.~(1) of the main text, has been obtained starting from its definition provided in Ref.~\cite{Beenakker2013} as the ground-state fermion parity number:
\begin{equation}
	\label{Pfaf_suppl}
	\sigma(\phi,\theta)=\sign{\left[\frac{\text{Pf}(r_As_0-s_0^Tr_A^T)}{\sqrt{\text{Det}[is_0]}}\right]}_{\epsilon=0}.
\end{equation}
Here $\text{Pf}(\cdot)$ is the Pfaffian of an antisymmetrized matrix built from $r_A$ and $s_0$, see in Eqs.~\eqref{s0_comb}-\eqref{sM_sL_sR}. According to Refs.~\cite{Fu2009a,Beenakker2013}, this figure of merit $\sigma = \pm$ can take two possible values corresponding to an even ($\sigma=+$) or odd ($\sigma=-$) number of electrons in the ground state of the system, and can change its value only at zero-energy crossings of the ABS spectrum. 
In our setup, it does not correspond to the parity of the ground state, but nevertheless allows us to single-out remarkable protection features of the zero-energy crossings of the IQH TJJ ABS spectrum. 
We computed the Pfaffian in Eq.~\eqref{Pfaf_suppl} in presence of ordinary reflections $\rho \neq 0$, showing analytically that it does not depend on $\rho$, see Eq.~(1) in the main text.
It would be interesting to understand in future works the validity and interpretation of $\sigma$ for more diverse setups. 

\begin{figure}[htb!]
	\centering
	\includegraphics[width=1\columnwidth]{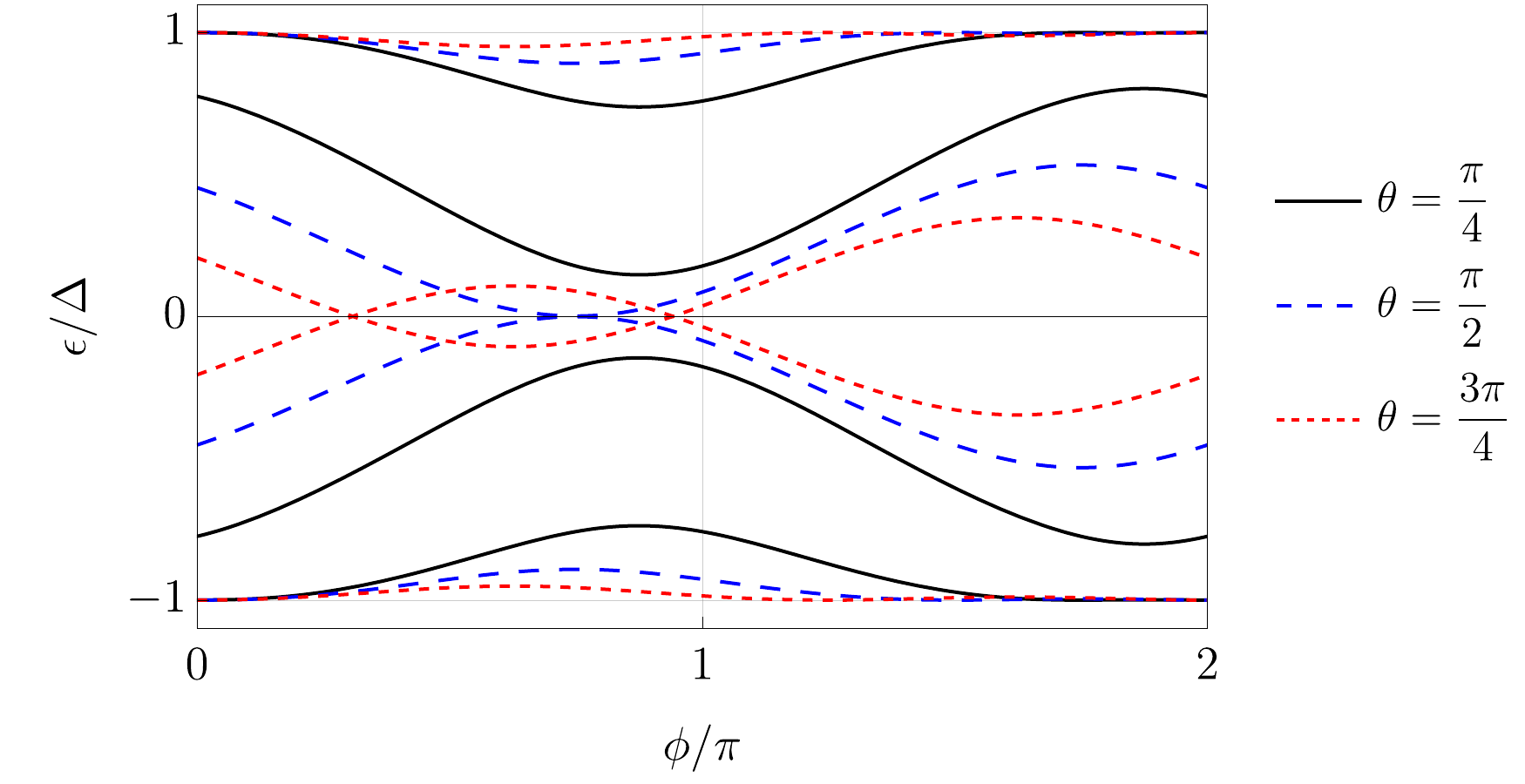}
	\caption{ABS spectrum in the case of an asymmetric junction when $r=0$ with $\rho_L=\sqrt{0.5}$ and $\rho_R=\sqrt{0.2}$ for different values of $\theta$. When $\theta=\frac{\pi}{4}$ (black solid curves) a zero energy gap is open. By changing $\theta$ the gap closes in one point for $\theta=\frac{\pi}{2}$ (blue dashed curves) and two points for $\theta=\frac{3\pi}{4}$ (red dotted curves).}
	\label{ABS_rho1_rho2}
\end{figure}

\section{Asymmetric junction}
\label{app_Asymmetric_junction}

In the main text we demonstrated that the occurrence of zero-energy crossings is protected as long as the system is left/right symmetric, i.~e.~when $\rho_L=\rho_R$. 
In particular, when $r=0$, a zero energy gap cannot be opened in the ABS spectrum for any value of $\theta$ (see left panels of Fig.~2 of the main text).
When the junction is asymmetric, i.e.~$\rho_L\neq \rho_R$, the situation radically differs as it becomes possible to open a gap even when $r=0$. We illustrate this in Fig.~\ref{ABS_rho1_rho2}. However, by tuning $\theta$ (for any choice of $\rho_L$ and $\rho_R$) it is possible to close the gap again.
In other words, by properly gating with $V_g$ the IQH TJJ, we can restore the protected crossing by \emph{effectively} compensating the right/left asymmetry. 
In Fig.~\ref{ABS_rho1_rho2}, we show the ABS spectrum of the junction when $r=0$ with $\rho_L=\sqrt{0.5}$ and $\rho_R=\sqrt{0.2}$ for different values of $\theta$ (similar results can be obtained when $\rho_L$ and $\rho_R$ are complex amplitudes with different modulus and phase). The black solid curves represent the ABS energies as a function of $\phi$ when $\theta=\frac{\pi}{4}$. 
In this case, a zero energy gap opens, i.e. no zero energy crossings occur. However by varying $\theta$, the lowest energy ABS moves closer to the zero energy axis until it crosses it in one point for $\theta=\frac{\pi}{2}$ (blue dashed curves), for this parameters, and two points for $\theta=\frac{3\pi}{4}$ (red dotted curves).
This behavior shows that, even if a zero energy gap opens for asymmetric junctions, the emergence of the crossing at zero energy can always be obtained by changing $\theta$ via the external gate voltage $V_g$.

\section{ABS excitation spectrum: Andreev qubit}
\label{app_Andreev_qbit}

\begin{figure}[htb]
	\centering
	\includegraphics[width=1\columnwidth]{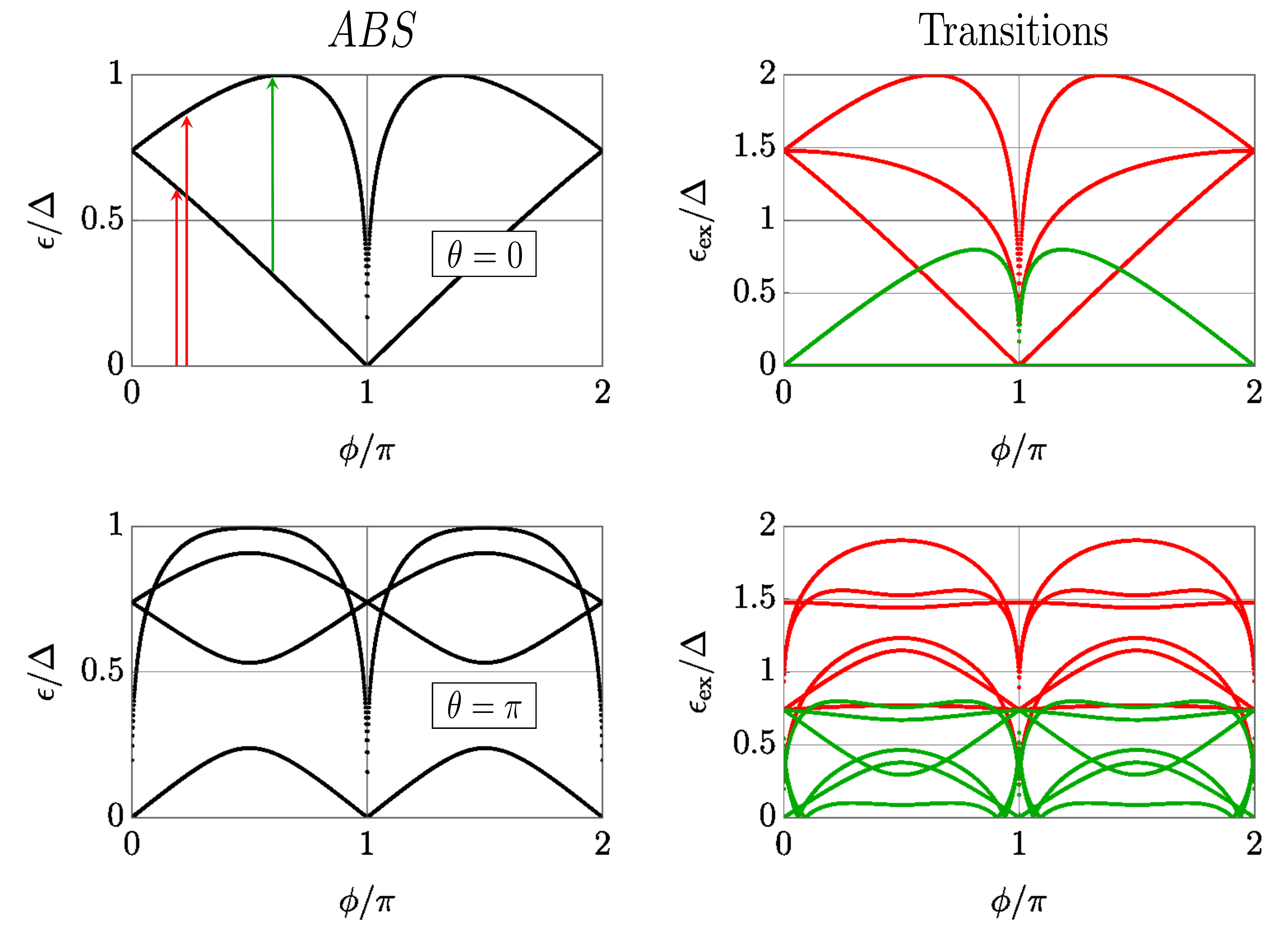}
	\caption{ABS spectrum (left panels) and ABS excitation spectrum (right panels) of the Andreev qubit for an IQH TJJ for two different values of $\theta=0$ (top panels) and $\theta=\pi$ (bottom panels). The red and green lines in the  ABS excitation spectrum correspond to two different types of processes: a Cooper pair is broken exciting two quasiparticles states in the ABS spectrum (red lines) or quasiparticles state transition between two different ABS states. Other parameters are $L=W=\xi$, $B=3~T$, $\abs{\rho}^2=0.1$ and $\abs{r}^2=0$.}
	\label{AndreevQbit}
\end{figure}

In order to characterize the IQH TJJ as a potential platform for implementing Andreev qubits, we investigate the ABS excitation spectrum that can be probed in the microwave range. 
% For this application, it is realistic to consider a finite-length weak-link ($L=\xi$) with non vanishing ordinary reflections at interfaces, i.e~$\abs{\rho}^2=0.1$.
% \blue{For this application, it is realistic to consider a finite-size weak-link ($L=W=\xi$) with non-vanishing ordinary reflections at the interfaces, i.e., $\abs{\rho}^2=0.1$. In this case, we consider a magnetic flux $\Phi=30~ \Phi_0$, corresponding to an applied magnetic field of $B=3~T$.}
For this application, it is realistic to consider a finite-size weak-link ($L=W=\xi$) with non-vanishing ordinary reflections at the interfaces, i.e., $\abs{\rho}^2=0.1$, and an applied magnetic field of $B=3~T$.
In Fig.~\ref{AndreevQbit}, for two values of $\theta$, we show the ABS spectrum (on the left column) and the ABS excitation spectrum (on the right column) as functions of the phase difference $\phi$. 
For the ABS excitation spectrum we distinguish between two different types of transition lines. We show with red arrows in the top-left panel the Cooper pair breaking processes where microwave absorption breaks a Cooper pair exciting two quasiparticles in the ABS spectrum. 
With green arrows in the top-left panel we denote the single-particle transitions where a trapped quasiparticle already occupying an Andreev state is excited to another one. 
For the two different classes, the transition energies of the excitation spectrum can be obtained by adding the energies of two different ABS levels (in the case of red transitions) or by subtracting them (in the case of green transitions).

The finite length of the weak-link introduces higher energy branches in the ABS spectrum. 
In particular, for $\theta=0$ (see upper left panel of Fig.~\Ref{AndreevQbit}), we see that the ABS spectrum presents two branches instead of just one as shown in the case of short junction -- see left panel of Fig.~1~$(c)$ of the main text.
Nevertheless, similarly to the short junction case, both branches of the ABS spectrum here exhibit a zero-energy protected crossing at $\phi=\pi$.
Notice also, that in this case the ABS spectrum is degenerate, and its signature can directly be seen in the ABS excitation spectrum (see upper right panel of Fig.~\Ref{AndreevQbit}): the breaking of a Cooper pair (red curves) can excite two degenerate states, hence allowing some of the red curves to be exactly twice the energy of the spectrum of the ABS.
\\
For $\theta=\pi$ (bottom panels), instead, we get four non-degenerate ABSs: two of them exhibit zero-energy protected crossings at $\phi=0,\pi$ (similarly to the short junction case as shown in the right panel of Fig.~1~$(c)$ of the main text), while two do not cross the zero-energy axis.
This behavior reflects in a richer structure of the ABS excitation spectrum as shown in the bottom right panel of Fig.~\Ref{AndreevQbit}. 
This confirms the high tunability of the IQH TJJ using electrostatic controls to be exploited as a potential platform for the implementation and manipulation of Andreev qubits.

%\nocite{*}
\bibliography{biblio}

\end{document}